\documentclass[mathpazo]{cicp}
\usepackage{amssymb,amsmath,bm,graphicx,subfigure,color}

\newcommand{\figref}[1]{Fig.~\ref{#1}}

\newcommand{\eqnref}[1]{Eq.~\eqref{#1}}

\newcommand{\pre}{Phys.~Rev.~E}
\newcommand{\jcp}{J.~Chem.~Phys.}
\newcommand{\prl}{Phys.~Rev.~Lett.}

\let\oldhat\hat

\renewcommand{\hat}[1]{\oldhat{\mathbf{#1}}}

\begin{document}
\title{Enslaved Phase-Separation Fronts and Liesegang Pattern Formation}
\author[Foard et.~al]{E.M.~Foard\corrauth, and A.J.~Wagner}
\address{Department of Physics, North Dakota State University, Fargo, ND 58105}
\emails{{\tt eric.foard@nds.edu} (E.M. Foard), {\tt alexander.wagner@ndsu.edu} (A.J.~Wagner)}

\begin{abstract}
We show that an enslaved phase-separation front moving with diffusive speeds $U=C/\sqrt{T}$ can leave alternating domains of increasing size in their wake.  We find the size and spacing of these domains is identical to Liesegang patterns.  For equal composition of the components we are able to predict the exact form of the pattern analytically.  We also show that there is a critical value for $C$ below which only two domains are formed.  Our analytical predictions are verified by numerical simulations using a lattice Boltzmann method.
\end{abstract}


\maketitle

\section{Introduction to Liesegang Patterns}\label{liesegang}
The formation of highly ordered patterns in naturally occurring biological, chemical, and mineralogical systems has long been a subject of intense interest.  The study of such pattern formation can sometimes allow deep insight into their underlying natural phenomena.  In a previous paper we analyzed the dynamics of pattern formation behind a one-dimensional, slow moving (enslaved), phase-separation front.  Our analysis concerned fronts moving with constant speed, and the pattern formed was a series of alternating bands of regular width and spacing\cite{foard-2009}.  We show here that a front moving with diffusive speed will form a more complex Liesegang pattern.

It was with the motivation of understanding pattern formation in simple systems that, just over one hundred years ago, R. E. Liesegang observed a highly ordered pattern of concentric rings precipitating around a drop of silver nitrate on a glass slide with a thin gel coating containing potassium dichromate\cite{liesegang-1896}.  These concentric rings are now known as Liesegang rings. The radial width and spacing of the rings increases with increasing distance from the center.  Rings close to the center are narrow and tightly packed.  Rings far from the center are wide and far apart.  The pattern forms from the center outwards, and is stationary once visible.  Several alternatives to silver nitrate and potassium dichromate in the production of Liesegang rings have been used in the literature.  In general, some electrolytes $A$ and $B$ combine to form an insoluble precipitate $D$ which then produces Liesegang patterns.  Most recent publications use linear Liesegang patterns of bands and gaps which are produced by adding the $A$ electrolyte to a test-tube containing the $B$ electrolyte suspended in gel\cite{racz-1999}.

To characterize Liesegang bands or rings, they are typically numbered from the first formed to the last formed, the $n$th appearing at time $t_n$ at position $x_n$ with a width $w_n$.  Repeated careful measurements of Liesegang patterns revealed that discrete, defect-free bands could be characterized by a set of empirical laws\cite{jablczynski-1923}.  They are
\begin{eqnarray}
	\label{time_law}\mbox{Time Law}&& x_n \propto \sqrt{t_n} \;, \\
	\label{spacing_law}\mbox{Spacing Law}&& x_{n+1}/x_n = 1+p \;, \\
	\label{width_law}\mbox{Width Law}&& x_n \propto w_n \;.
\end{eqnarray}
The \emph{time law} relates the position of the $n$th band with the time of its appearance.  The location of subsequent bands is given by the \emph{spacing law}, where $p>0$ is the spacing coefficient.  The \emph{width law} states that the band width is proportional to the position of the band, which is a natural result of the \emph{spacing law} with the assumptions of mass conservation and uniform concentration of precipitate bands\cite{jahnke-2008}.  These laws are only considered valid for \emph{large $n$}.
Much attention has been paid to the phenomenological Matalon-Packter law $p = F(B_0) + G(B_0)/A_0$, which relates $p$ to the initial concentrations $A_0$ and $B_0$ of the $A$ and $B$ electrolytes\cite{antal-1998}.  $F(B_0)$ and $G(B_0)$ are known to be decreasing functions of $B_0$, though not much else about them is known.  Antal \textit{et al.}\cite{antal-1998} show that the Matalon-Packter law can be derived in limiting cases from more general expressions.

Soon after their characterization, study into the nature and cause of Liesegang patterns was prolific.  For instance, a paper by Stern in 1954 makes mention of more than six hundred papers having been published on the subject by that time\cite{stern-1954}.  There were many early attempts to develop a comprehensive model of Liesegang pattern formation.  However, it proved difficult to account for the wide variety and complexity of possible patterns, and many early models were eliminated by additional experiments.  The complexity of Liesegang pattern formation thwarted theoretical understanding, and progress slowed.  More than a century later there is currently still no generally accepted comprehensive mechanism for Liesegang pattern formation.

Current dominating theories of Liesegang pattern formation can be categorized as either an \emph{ion-product supersaturation} theory where electrolytes combine directly into the precipitate ($A + B \rightarrow D$), or a \emph{nucleation and growth} theory where one or more intermediate compounds form before final precipitation ($A + B \rightarrow C \rightarrow D$).  A brief discussion of these models is given in the recent paper by Jahnke and Kantelhardt\cite{jahnke-2008}.

For either of these theories, precipitation occurs behind a reaction front formed by the $A$ electrolyte diffusing into the $B$ electrolyte.  Since the initial concentration ($A_0$) of $A$ in the drop is typically an order of magnitude higher than the concentration ($B_0$) of $B$ in the gel, the time dependent concentration profile $A(x,t)$ of $A$ electrolyte in the gel forms a reaction front which resembles the familiar heat-diffusion profile moving with speed $u_r\propto t^{-1/2}$\cite{antal-1998}.  Coincident with, or trailing behind this reaction front is a precipitation front, moving at a speed $u_p\propto t^{-1/2}$, which produces the banded pattern.  The precise relationship of the precipitation front to the reaction front depends on the specific chemical and physical mechanism of Liesegang pattern formation which are still not entirely understood.

The recent trend for publications on Liesegang patterns shows a re-surging interest.  A search of the ISI Web of Science Internet database for publications on the topic ``Liesegang'' shows an increase in papers since a fall in the early second half of the 20th century.  We found only 8 papers from 1970 to 1979, 35 from 1980 to 1989, 107 from 1990 to 1999, and 178 since 2000.  Much of the recent research into Liesegang pattern formation has focused on the moving front by employing a variety of numerical techniques to simulate models of precipitation fronts.  Some examples are the reaction-diffusion cellular automata simulations by Chopard \textit{et al.}\cite{chopard-1994}, direct simulation of a model-B system with a chemical reaction like source term by Antal \textit{et al.}\cite{antal-1999} and R\'acz\cite{racz-1999}, the discrete stochastic simulation which used random walkers to model the diffusion front by Izs\'ak and Lagzi \cite{izsak-2003}, as well as the lattice gas simulation by Jahnke and Kantelhardt\cite{jahnke-2008}.

In this paper we show that patterns identical to Liesegang patterns can be formed in a much simpler physical system.  We consider a binary mixture that can phase-separate if a control parameter crosses a critical value.  In this system an enslaved phase-separation moves at a speed of $u = c t^{-1/2}$.  An example of such a system would be a binary mixture that is cooled below its critical temperature from one end.  This situation is somewhat similar to that of the electrolytes, where material is formed at the front and will subsequently phase separate.  However the details are quite different, most notably in the nucleation conditions.

It should be noted that this model is somewhat similar to the Model-B precipitation front proposed by Antal \textit{et al.}\cite{antal-1999}, but there are several key differences.  Their model has a moving, Gaussian shaped source of $\mathcal{A}$-type material designed to mimic the product of a chemical reaction front.  Their source moves through a region which is phase-separated into an equilibrium $\mathcal{B}$-type.  When the concentration of $\mathcal{A}$-type material in a given area reaches the spinodal value, an $\mathcal{A}$-type domain nucleates and depletes the surrounding region of its $\mathcal{A}$-type material.  The source moves on, leaving stable domains.  The speed, width, and concentration of their source are free parameters of their model.  Our model has an abrupt control parameter front which induces phase-separation as it passes through a mixed material, the mechanism for nucleation of new domains is quite different.  Notably switching does not occur at the spinodal value for reasons explained in\cite{foard-2009}.  Also the analysis of Antal \textit{et al.} is numerical in nature.  Our model has only one parameter and can be solved analytically.

To allow for an analytical treatment we make the simplifying assumption of an abrupt front.  We expect that the results will be qualitatively similar to those for an extended control parameter front.  In practice, an abrupt front could be experimentally achieved by immersing a thin sample with a prescribed speed $u\propto t^{-1/2}$ into a temperature bath.

The key result is that we are able to analytically determine the resulting patterns and we show that they obey the Liesegang laws given by Equations \eqref{time_law}, \eqref{spacing_law} and \eqref{width_law}.  We derive an analytical expression for the spacing coefficient $p$ in terms of the free parameters of this model.  We verify these theoretical predictions by direct numerical simulations using a lattice Boltzmann method.

\section{A Model for Liesegang Patterns Formed by Enslaved Phase Separation Fronts}
We consider two materials, an $\mathcal{A}$-type and a $\mathcal{B}$-type, in an incompressible mixture such that the total density $\rho = \rho_\mathcal{A}(x,t) + \rho_\mathcal{B}(x,t)$ remains constant.  The relevant variable is then the relative concentration of $\mathcal{A}$ to $\mathcal{B}$-type material defined as
\begin{equation}
	\phi(x,t) = \frac{\rho_\mathcal{A}(x,t) - \rho_\mathcal{B}(x,t)}{\rho}\;.
\end{equation}
From here on the time and position dependence of the concentration will be implied.  For simplicity we assume that the two materials have the mixing free energy described by a $\phi^4$ law
\begin{equation}
	\label{eqn:free-energy}
	F=\int\!\!dx \left[ \frac{a(x,t)}{2}\phi^2 + \frac{b(x,t)}{4}\phi^4 + \frac{\kappa(x,t)}{2}(\nabla\phi)^2 \right] \;.
\end{equation}
The time and position dependence of the control parameters ($a$, $b$ and $\kappa$) are such that they constitute a spatially abrupt transition from the mixing region to the phase-separating region of the phase diagram.  For example:
\begin{equation}
	a(x,t) = a_S + (a_M - a_S) \Theta\left[\int u \,dt + x_0 \right] \;,\\
	\label{eqn:parameter}
\end{equation}
where 
$\Theta$ is the Heaviside step function, and the transition takes the form of a front moving with velocity $u(t)$.  The free energy of \eqnref{eqn:free-energy} has a single minimum for $a>0$, resulting in material \emph{mixing}.  When $a<0$ there are two minimums, resulting in the \emph{separating} of material.  The control parameters for the \emph{mixing} and \emph{separating} regions are denoted by subscripts $M$ and $S$ respectively.  The other parameters are defined similarly.

Since we assume incompressibility and a one-dimensional geometry of the system, hydrodynamics can be neglected here, and the dynamics is therefore purely diffusive
\begin{equation}
	\label{eqn:dynamics}
	\partial_t \phi  = \nabla \left[ m(x,t) \nabla \mu(x,t) \right]\;.
\end{equation}
The chemical potential is derived from the free energy as,
\begin{equation}
	\mu(x,t)=\frac{\delta F}{\delta \phi} = a(x,t) \phi + b(x,t) \phi^3 - \kappa(x,t) \nabla^2\phi \;,
\end{equation}
and the diffusive mobility $m$ is one of the control parameters which can vary across the front
similarly to \eqref{eqn:parameter}.  Aside from time and space dependence of the control parameters, this is the familiar Model-B.

This model for a moving front has many parameters which can affect the dynamics of phase separation, however not all of these parameters are independent.  As we elaborate on in our previous paper, introduction of appropriate time, space, and concentration scaling can non-dimensionalize the equations of motion\cite{foard-2009}.  For the remainder of this paper we will work entirely in the non-dimensional scales
\begin{equation}
		\label{eqn:nondimension}
	T=\frac{t}{t_{sp}} \;,\;\; X=\frac{x}{\lambda_{sp}} \;,\;\; \Phi=\frac{\phi}{\phi_{eq}}\;.
\end{equation}
Here $t_{sp}=4\kappa/ma^2$ and $\lambda_{sp}=2\pi\sqrt{-2\kappa/a}$ are the characteristic time and length scales of spinodal decomposition, and $\phi_{eq}=\sqrt{-a/b}$ is the positive equilibrium concentration of the phase-separated material.  Non-dimensional quantities will be denoted by capital letters.

Nondimensionalization reduces the free parameters to an independent set of four parameters: $U=u/u_{sp}$ is the speed of the front scaled by the natural speed of spinodal decomposition $u_{sp}=\lambda_{sp}/t_{sp}$, $M=m_M/m_S$ is the ratio of the diffusive mobility ahead of the front to behind the front, $A=-a_M/a_S$ is the depth of the quench into the unstable region of the phase diagram, and $\Phi_{in}=\phi_{in}/\phi_{eq}$ is the non-dimensional initial concentration of the mixed material.  As we have previously shown, the dynamics becomes particularly simple if we let the diffusive mobility ahead of the front be negligible $m_M\rightarrow0$.  In this case the dynamics ahead of the front, represented by \eqnref{eqn:dynamics}, is halted.  This makes the $A$ parameter unnecessary.  This is not a particularly physical assumption, but it does only cause a small quantitative change in the pattern formation as compares to $M=1$, as we have previously shown\cite{foard-2009}.  The existence of analytical solutions, however, makes $M=0$ an attractive choice.  If we then assume that the front moves into material which has an equally mixed initial concentration, the morphology of domains formed depends only on the non-dimensional front speed ($U$).  Interestingly, an analytical expression can be found for the non-dimensional domain wavelength ($L$) when $U$ is small and constant:
\begin{equation}
	L(U) = \frac{4\left( \sqrt{6} + 6\ln\left( 2-\sqrt{2/3} \right) -3 \right)}{3 \pi^2 U} = \frac{\Psi_0}{U} \;.
	\label{eqn:solution}
\end{equation}
The derivation of this law is given in\cite{foard-2009}.  Because it is convenient to work with this analytical solution, we will focus on the condition where the initial mixed material contains equal parts of $\mathcal{A}$-type and $\mathcal{B}$-type material.

To complete our model we require that the front moves with a time dependent speed
\begin{equation}
	U(T) = \frac{C}{\sqrt{T}}\;,
	\label{liesegang_speed}
\end{equation}
appropriate for a diffusive velocity.  The parameter $C$ now becomes the only free parameter in this model.  This parameter will therefore determine the spacing and width of the resulting domains.

\section{Derivation of Liesegang Laws}
We will now show that for certain parameter choices the model presented in the previous chapter will result in the formation of Liesegang patterns.  We will do this by deriving the Liesegang laws of Equations \eqref{time_law}, \eqref{spacing_law} and \eqref{width_law} directly from the model.  We will conclude this section by deriving a Matalon-Packter like analytical expression for the spacing coefficient.

We first recognize that domain production at any point in space can only occur after the phase-separation front has passed.  From our previous paper we observe that once an enslaved front has passed a point, domain growth or nucleation occurs very rapidly\cite{foard-2009}.  That is, domains form and grow near the front.  The position of the front at time $T$ is found by integrating the front-speed:
\begin{equation}
	\label{enslaved-time-law}
	X(T) = \int_0^T U(T)\,dT=\int_0^T \frac{C}{\sqrt{T}}\,dT = 2 C \sqrt{T} = \alpha \sqrt{T}\;.
\end{equation}
From this we see that if domains are formed, then the $n$th domain will be formed at time $T_n$ at position $X_n$ which is proportional to $\sqrt{T_n}$ in agreement with the time law of \eqref{time_law}.  This is simply a result of the imposed front speed.

Deriving the width and spacing laws is more interesting.  As we found in our previous paper\cite{foard-2009} and reproduced in \eqnref{eqn:solution}, a front moving at constant speed produces domains of a predictable wavelength.  In the process of deriving \eqnref{eqn:solution}, we discovered domain growth has two distinct stages: first is a formation stage where a domain nucleates at the front and grows as it is pulled along with the front until a new domain nucleates; second is an expansion stage where the just detached domain grows due to deposition of material excluded from the newly forming, opposite type domain.  For a constant speed front moving into material which is initially symmetrically mixed and of negligible mobility, each stage accounts for growing half of the domain's final width.  In essence, it takes two domain-type switching cycles to completely form a stable domain, and the domain width grows at half the front speed.

\begin{figure}
	\begin{center}
		\includegraphics[clip=true, scale=0.333333]{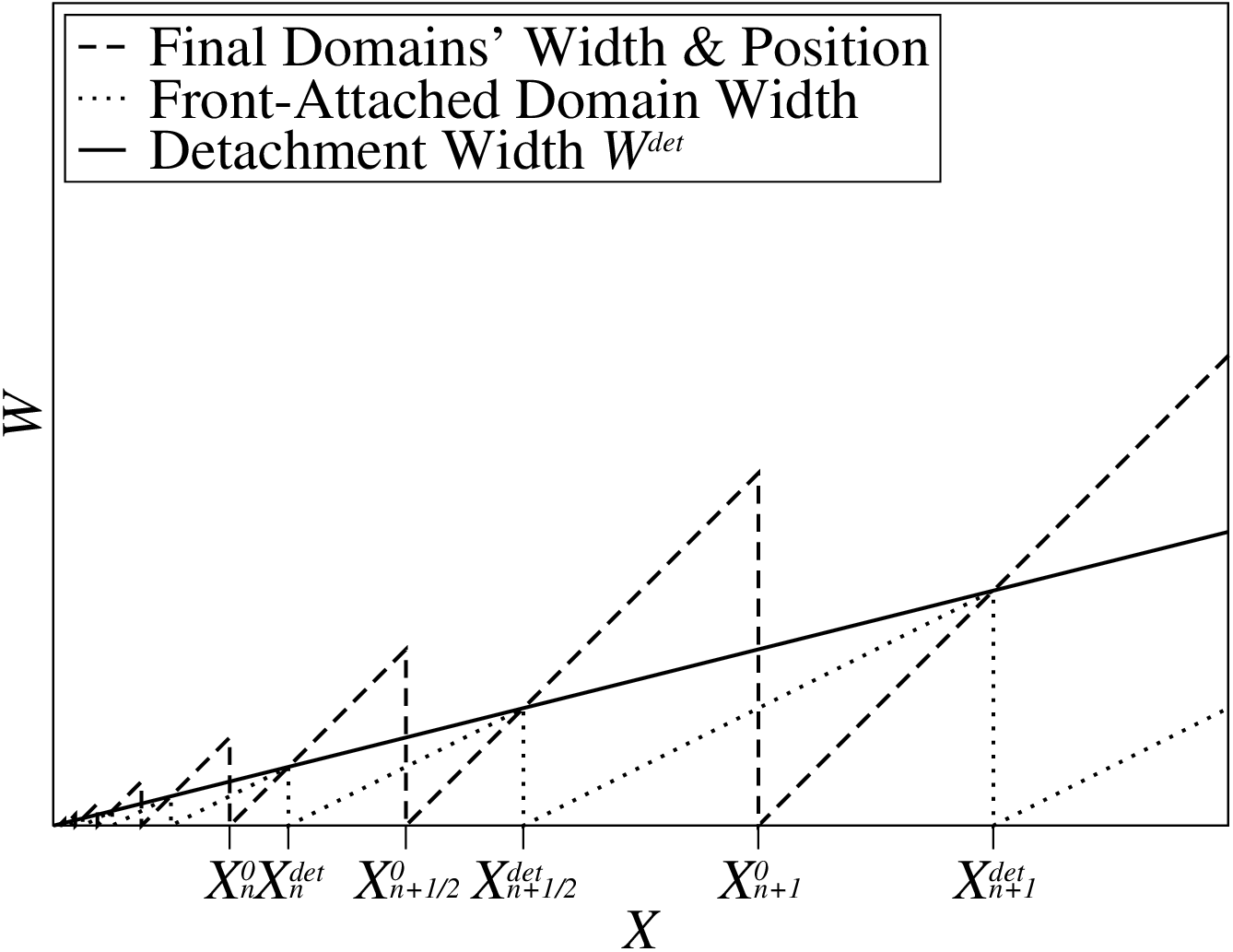}
		\caption{
			This figure shows three important aspects of Liesegang pattern formation.
			The dashed line represents the distance from the previous domain in final morphology.  We label the position of the first interface of a domain $n$ as $X_n^0$.  Therefore the dashed line is $X - X_n^0$ where $n$ is the domain that exists at position $X$.  The dashed line has slope $1$ because the $W$ and $X$ axes have the same scale.
			The dotted line represents the width $W$ of the front-attached domain as a function of the front position $X$.  The slope $1/2$ is explained in the text and in our previous paper on enslaved front phase-separation\cite{foard-2009}.
			The solid line represents the maximum front-attached domain width of \eqnref{detach_width} for the position $X$ of the front.  When a domain forming at the front reaches $W^{det}(X)$, it detaches from the front, and a new domain is nucleated at the front.
			The dotted, dashed, and solid lines all intersect at the points where domain type switching occurs at the front.  Note that when a domain $n$ becomes detached from the front its first interface becomes stationary at $X_n^0$.
		}
		\label{fig:growth}
	\end{center}
\end{figure}

The width of a domain as it detaches from a constant speed front is half of its final width, which is then a quarter of the constant speed domain wavelength for a front at that speed.  If the front speed is changing, the width at detachment $W^{det}$ is one quarter the domain wavelength predicted in \eqnref{eqn:solution} for a front moving with instantaneous speed $U$ at detachment.  This statement implies the assumption that our nucleation theory that was derived for constant $U$ can also be applied for time-dependent $U$.  This is a non-trivial assumption but it is justifiable be the excellent agreement of our theory with direct numerical simulations of the full PDE.
Using \eqnref{liesegang_speed} to replace the front speed dependence with time, then using \eqnref{enslaved-time-law} to replace the time dependence with position, we can predict the width $W^{det}$ of a domain as it detaches from a front when this front is at position $X$:
\begin{equation}
	\label{detach_width}
	W^{det} = \frac{1}{4}L(U)  = \frac{\Psi_0}{4U} = \frac{\Psi_0\sqrt{T}}{4C} = \frac{\Psi_0}{8C^2}X = \beta X\;.
\end{equation}
Note that this already resembles the Liesegang width law, but does not relate the final domain width to the domain position.  For that we must consider the growth of the domain after it detaches from the front.  The information presented in this section thus far is graphically shown in \figref{fig:growth} which represents the growth and final morphology of Liesegang pattern formation.

To derive the width law, recall the formation and expansion stages of domain growth mentioned previously.  A domain forming at the front increases its width at half the front speed.  Due to mass conservation, the detached domain directly behind it increases its width at the same rate.  The width of domains at detachment is predicted by \eqnref{detach_width}.  The final width of domain $n$ is then its detachment width, plus the detachment width of the next domain of opposite material, which we count as $n+1/2$.  From this we get:
\begin{equation}
	\label{width_1}
	W_{n} = W^{det}_{n} + W^{det}_{n+1/2} = \beta X^{det}_{n} + \beta X^{det}_{n+1/2}\;.
\end{equation}
The position of the first interface $X^{0}_{n}$ of domain $n$ can be easily found from the detachment position $X^{det}_{n}$ by subtracting the width $W^{det}_{n}$ of the domain at detachment:
\begin{equation}
	\label{width_2}
	X^{0}_{n} = X^{det}_{n} - W^{det}_{n} = \left(1 - \beta \right) X^{det}_{n}\;.
\end{equation}
As evident in \figref{fig:growth}, the position $X^{0}_{n+1/2}$ of the first interface of the next domain, which will be of opposite composition, is simply the position of the current domain's second interface, and is found by adding the current domain's width to its first interface position:
\begin{equation}
	\label{width_3}
	X^{0}_{n+1/2} = X^{0}_{n} + W_{n}\;.
\end{equation}
Combining Equations \eqref{width_1}, \eqref{width_2}, and \eqref{width_3}, we recover the Liesegang width law:
\begin{equation}
	\label{enslaved-width-law}
	W_n = \frac{1}{1/2\beta - 1} X^{0}_{n} = \frac{1}{4C^2/\Psi_0 - 1} X^{0}_{n} = \gamma X^{0}_{n}\;.
\end{equation}
We now derive the spacing law.  As seen from \eqnref{width_3}, the position of the first interface of the subsequent domain scales by a constant factor:
\begin{equation}
	X^{0}_{n+1/2} = X^{0}_{n} + W_n = (1 + \gamma)X^{0}_{n}\;.
\end{equation}
Two subsequent domains increase the index $n$ by one, and their positions scale by that same factor squared:
\begin{equation}
	X^{0}_{n+1} = (1 + \gamma)X^{0}_{n+1/2} = (1 + \gamma)^2X^{0}_{n}\;.
\end{equation}
This gives the Liesegang spacing law
\begin{equation}
	\label{enslaved-spacing-law}
	\frac{X^{0}_{n+1}}{X^{0}_{n}} = (1 + \gamma)^2 = \frac{1}{\left( 2\beta - 1 \right)^2} = \frac{1}{ \left( \Psi_0/4C^2 - 1 \right)^2}\;.
\end{equation}
The position of the band is taken as the boundary between the band and the previous gap.  Due to experimental reasons, the location of a Liesegang band is often considered to be at the center of the band, \textit{i.e.} $X_{n}=X^{0}_{n} + W_{n}/2$.  Our derivation of the Liesegang time, spacing, and width laws uses the fact that there is some freedom in measuring the positions of the domains $X_{n}$, as long as all domains are measured consistently.  It is clear that the scaling law holds whether one measures the position on the leading or trailing edge of the domain edge.  Note that if one definition of the position given by
\begin{equation}
	\label{eqn:generic_position}
	X_{n} = X^{0}_{n} + \nu W_{n}\;,
\end{equation}
obeys the Liesegang laws, then all these definitions of the position will still fulfill the Liesegang spacing laws because:
\begin{equation}
	\frac{X_{n+1}}{X_{n}}
	= \frac{X^{0}_{n+1} + \nu W_{n+1}}{X^{0}_{n} + \nu W_{n}}
	= \frac{ \left( 1+\gamma\nu \right) X^{0}_{n+1}}{\left( 1+\gamma\nu \right) X^{0}_{n} }
	= \frac{X^{0}_{n+1}}{X^{0}_{n}}\;.
\end{equation}

From \eqnref{enslaved-spacing-law}, we determine that the Liesegang spacing coefficient for this model is:
\begin{equation}
	\label{enslaved_matalon}
	p = \frac{1}{\left( \Psi_0/4C^2 - 1 \right)^2} - 1\;.
\end{equation}
Note that this expression replaces the Matalon-Packter law, but bears no apparent resemblance to this phenomenological law.  For values of $C>\sqrt{\Psi_0}/2$, the Liesegang spacing coefficient $p$ is greater than zero, and many domains will form, generating a Liesegang pattern; $C^{cr} = \sqrt{\Psi_0}/2 \approx 0.1247$ is a critical value.  For $C$ below this value no Liesegang patterns will be formed.  Instead only two domains continue to grow and no nucleation of new domains occurs.  A graphical representation of \eqnref{enslaved_matalon} is shown in \figref{fig:p_theory}.  The authors are not aware of any previous purely analytical expressions that accurately predict the spacing coefficient for any Liesegang pattern producing models.

We have now derived all of the laws for Liesegang patterns.  Surprisingly there are no free parameters and our analytical expressions completely determine the patterns.  It remains to show that the approximations made in this analytical derivation do not significantly alter the results.  To do this we compare our analytical results to direct numerical simulations of our model.

\section{Numerical Method and Results}
In our previous paper we present a one-dimensional lattice Boltzmann method (LBM) simulation of our model for enslaved phase-separation fronts moving with constant speed.  We did this by creating a lattice with spatially dependent control parameters, where the parameters change abruptly at the location of the front.  We then advect the material at speed $u$ across this stationary front, using carefully constructed inflow and outflow boundary conditions.  A Galilean transformation $x'\rightarrow x-ut$ of this simulation returns the original model.  The simulation was designed this way, because a slow moving abrupt front on a discrete spatial lattice would be stationary for long periods followed by an instantaneous change of position. On the other hand, since the current of material in a LBM simulation is represented by continuous distribution functions, a constant drift speed $u$ in the drift diffusion of material can be made arbitrarily small.  For details on the development of the LBM simulations for enslaved fronts please refer to our previous paper\cite{foard-2009}.  Here we will only discuss the changes required to implement this method for fronts moving with a non-constant speed.

\begin{figure}
	\begin{center}
		\subfigure[]{\label{fig:profile}\includegraphics[clip=true, scale=0.3]{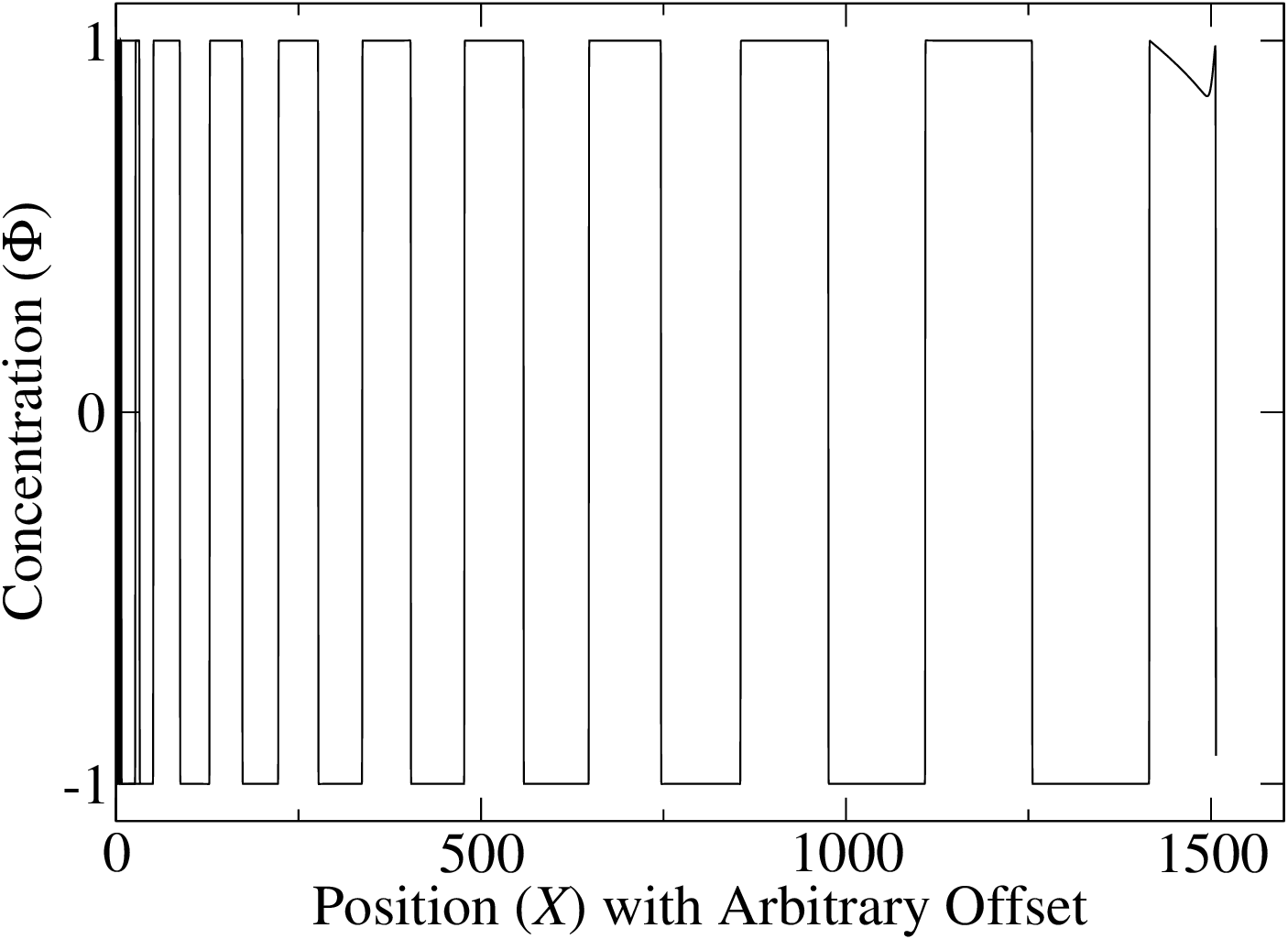}}
		\hfill
		\subfigure[]{\label{fig:fitting}\includegraphics[clip=true, scale=0.3]{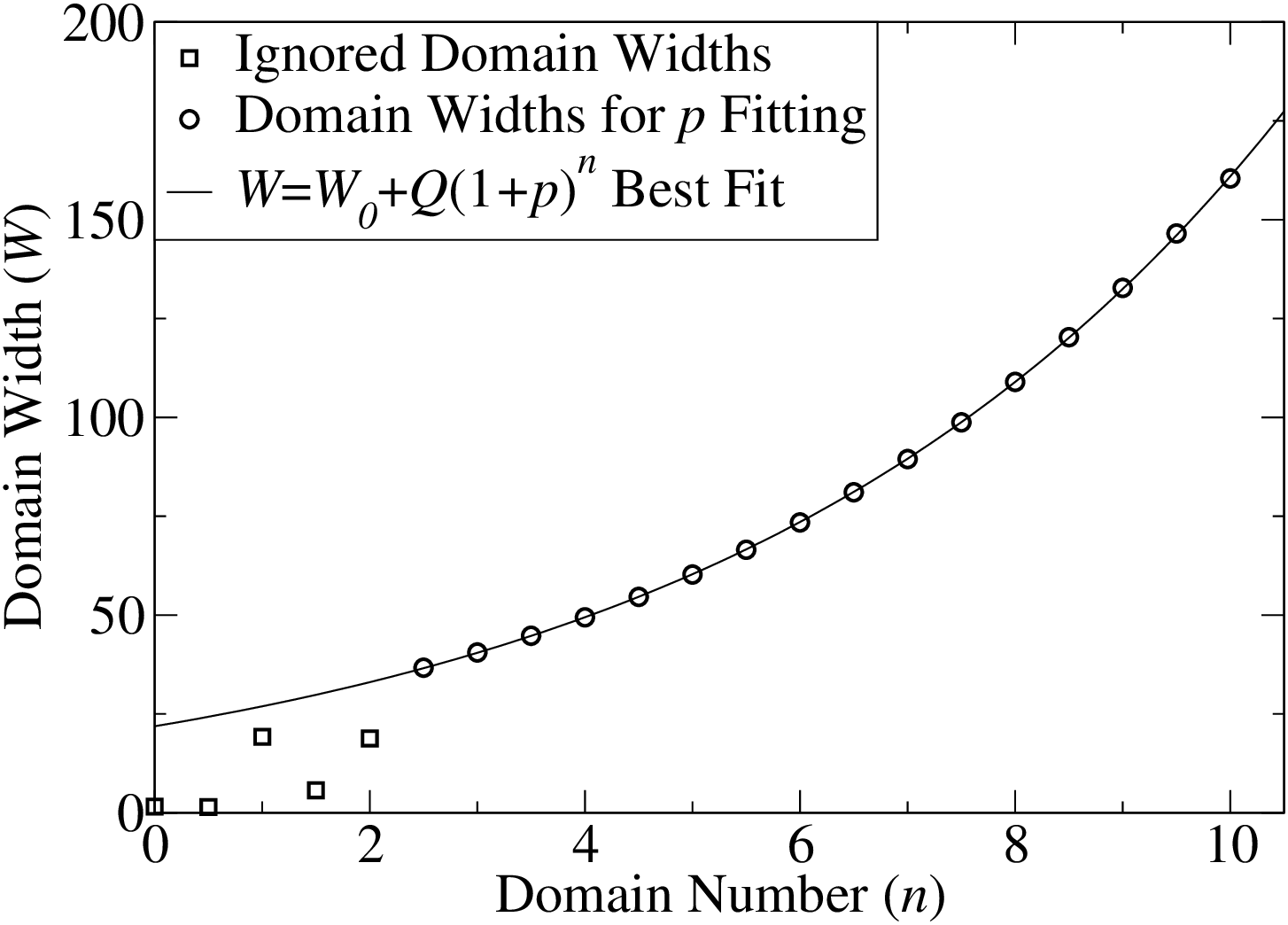}}
		\caption{
		Final data from LBM simulation for $C=0.4$.
		Subfigure (a) shows an example concentration profile of Liesegang pattern morphology at simulation completion.  The offset is due to requiring an initial finite speed and dynamic growing of the simulation.
		Subfigure (b) shows the fitting the domains widths found in (a) to determine $p(C)$.  Some domains are ignored because they are not formed by the front, but are instead artifacts of the simulation procedure.
		The width $W_n$ of domain $n$ is fitted to the equation $W_n = W_0 + Q(1 + p)^n$ to find an experimental value of $p$ to compare to the theoretical prediction of \eqnref{enslaved_matalon}.  For these data the fitting values are $W_0=-1.92468$, $Q=23.7999$, and $p=0.212093$.}
		\label{fig:sims}
	\end{center}
\end{figure}

\subsection{Changes for Diffusive Fronts and Details of Implementation}
The method we presented was designed with constant speed fronts in mind, but there was no requirement made that the front speed, and therefore the material advection speed in the LBM simulation, be constant.  To implement our model for the diffusive speed front of \eqnref{liesegang_speed} we make two changes.  First, the simulation is started at time $T=T_0>0$ to ensure that the advection speed is finite.  Second, the advection speed is recalculated according to \eqnref{liesegang_speed} at every iteration.  To ensure the analytical solution of \eqnref{eqn:solution} is applicable, we must use front speeds that are sufficiently slow.  The numerical verification of \eqnref{eqn:solution} is given in \cite[Fig. 7]{foard-2009} and shows that fronts moving at non-dimensional speed of $U\le0.001$ produce domain sizes that agree very well with the prediction.  This being the case, we begin our simulation so that the initial front speed is $U_0=0.001$, by setting our simulation start time $T_0=(C/U_0)^2$.

The simulation is initialized with random concentration fluctuations around the symmetrically mixed concentration value of $\Phi_{in}=0$.  This results in spinodal decomposition in the region behind the front.  These small domains serve to buffer the domains formed by the enslaved front from being adversely effected by any anomalies due to the outflow boundary condition.  To increase simulation performance the number of lattice sites is initially rather small, and is dynamically grown to maintain a buffer of small domains.

\subsection{Measurements and Calculations}
\begin{figure}
	\begin{center}
		\includegraphics[clip=true, scale=0.333333]{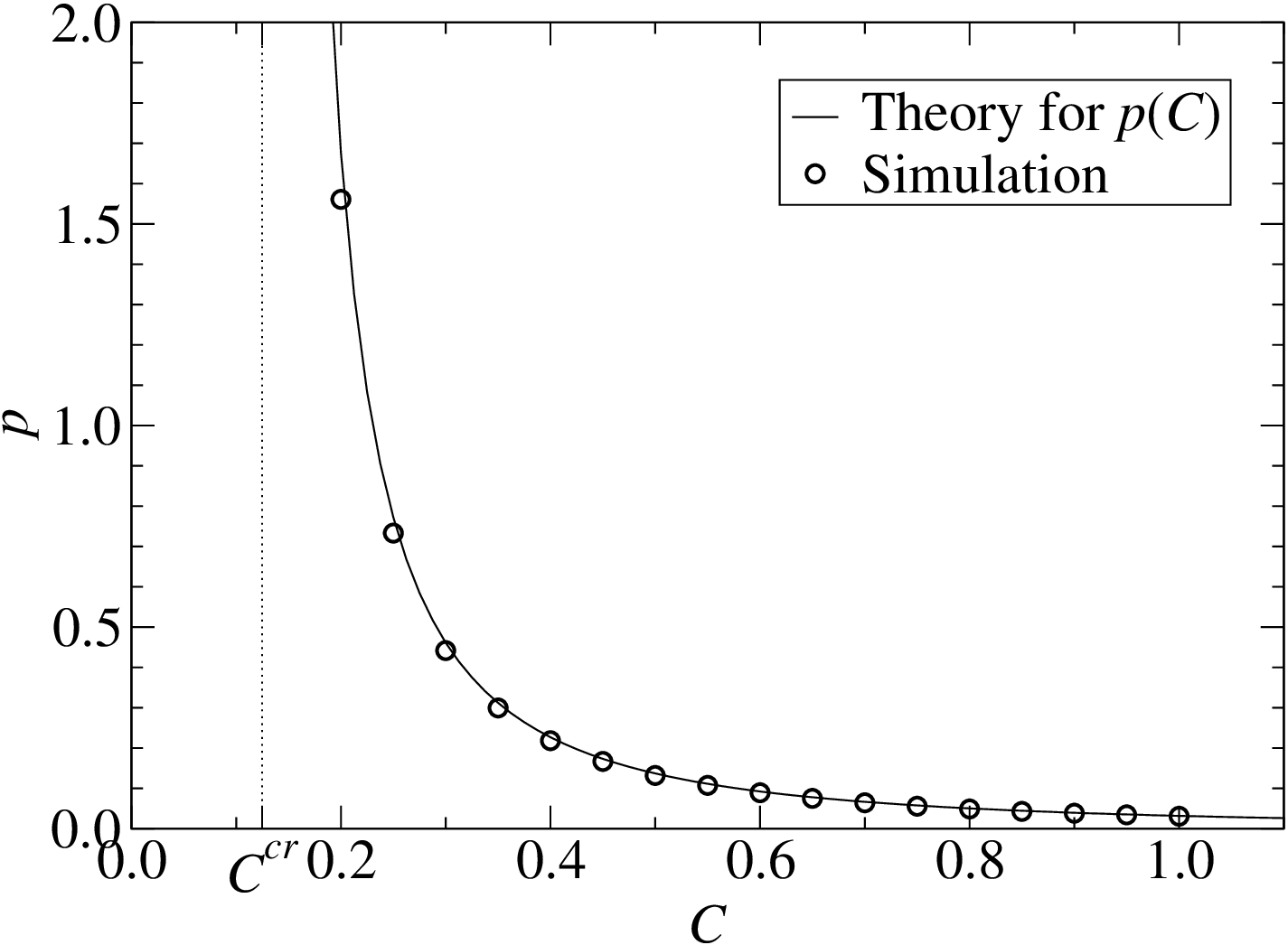}
		\caption{Comparison of the theoretical prediction for $p$ as a function of the free parameter $C$ shown in \eqnref{enslaved_matalon}.  This expression replaces the Matalon-Packter law for this model.  The results show excellent agreement of the simulation results with the assumptions made in the derivation of the Liesegang scaling laws for this model.  For $C<C^{cr}$ no new domains were nucleated.  These data were taken from a series of simulations, the details of which are explained in the text.}
		\label{fig:p_theory}
	\end{center}
\end{figure}

Our goal with these simulations is to verify our analytical predictions of Equations \eqref{enslaved-width-law} and \eqref{enslaved-spacing-law} for the Liesegang laws.  We accomplish this by performing simulations with different values of $C$, measuring the width $W_n$ of the $n$th domain formed, and performing a numerical fitting to the equation
\begin{equation}
	\label{simulation-fitting}
	W_n = W_0 + Q\left( 1 + p \right)^n\;,
\end{equation}
where $W_0$, $Q$ and $p$ are the fitting parameters.  This is an alternative form of \eqnref{spacing_law} with a substitution provided by \eqnref{width_law}.

The domain widths are measured by interpolating the sub-lattice position of the zero-crossing of the concentration at an interface, and determining the distance to the sub-lattice position of the next interface.  To ensure that there are sufficient domains to perform meaningful fitting to \eqnref{simulation-fitting} we track the number of switching events; after 19 switching events have occurred the domain widths and other data, such as the concentration profile, are written out to disk.  As explained in the previous section, the two domains directly behind the front are not completely formed, and are not used to find the experimental $p$ value.  The first domain formed by the front may also be ignored as it can sometimes be induced by the very strong dynamics of spinodal decomposition.  This typically leaves 16 alternating domains of $\mathcal{A}$ and $\mathcal{B}$-type material which can be used for fitting to \eqnref{simulation-fitting}.  Example simulation data and $p$ fitting are shown in \figref{fig:sims} for the $C=0.4$ data point shown in \figref{fig:p_theory}.  The concentration profile is shown in \figref{fig:profile}, and the width fitting is shown in \figref{fig:fitting}.  

One additional note: the concentration profile of \figref{fig:profile} seems to show a very sharp interface.  The actual interface width covers approximately ten lattice sites, and is in agreement with the bulk stability requirements outlined by Wagner and Pooley in \cite{wagner-2007}.  The use of the minimum stable interface width increases simulation spatial efficiency.  Additionally, the use of smaller interface widths allows a larger time scaling factor $s$ to be used which further increases simulation efficiency by calculating a larger time step each iteration.

Numerous simulations of the type shown in \figref{fig:sims} were performed for different values of $C$ and the corresponding $p$ values were measured and compared to the theoretical prediction of \eqnref{enslaved_matalon}.  These results are shown in \figref{fig:p_theory} and show excellent agreement.  This remarkable agreement suggests that approximations made in the derivation of the analytical results have negligible effect on the final results.  This is notable, as no other Liesegang pattern forming model has an accurate analytical prediction of the Liesegang scaling laws.

\section{Outlook}
\begin{figure}
	\begin{center}
		\subfigure[]{\label{fig:off_profile}\includegraphics[clip=true, scale=0.29]{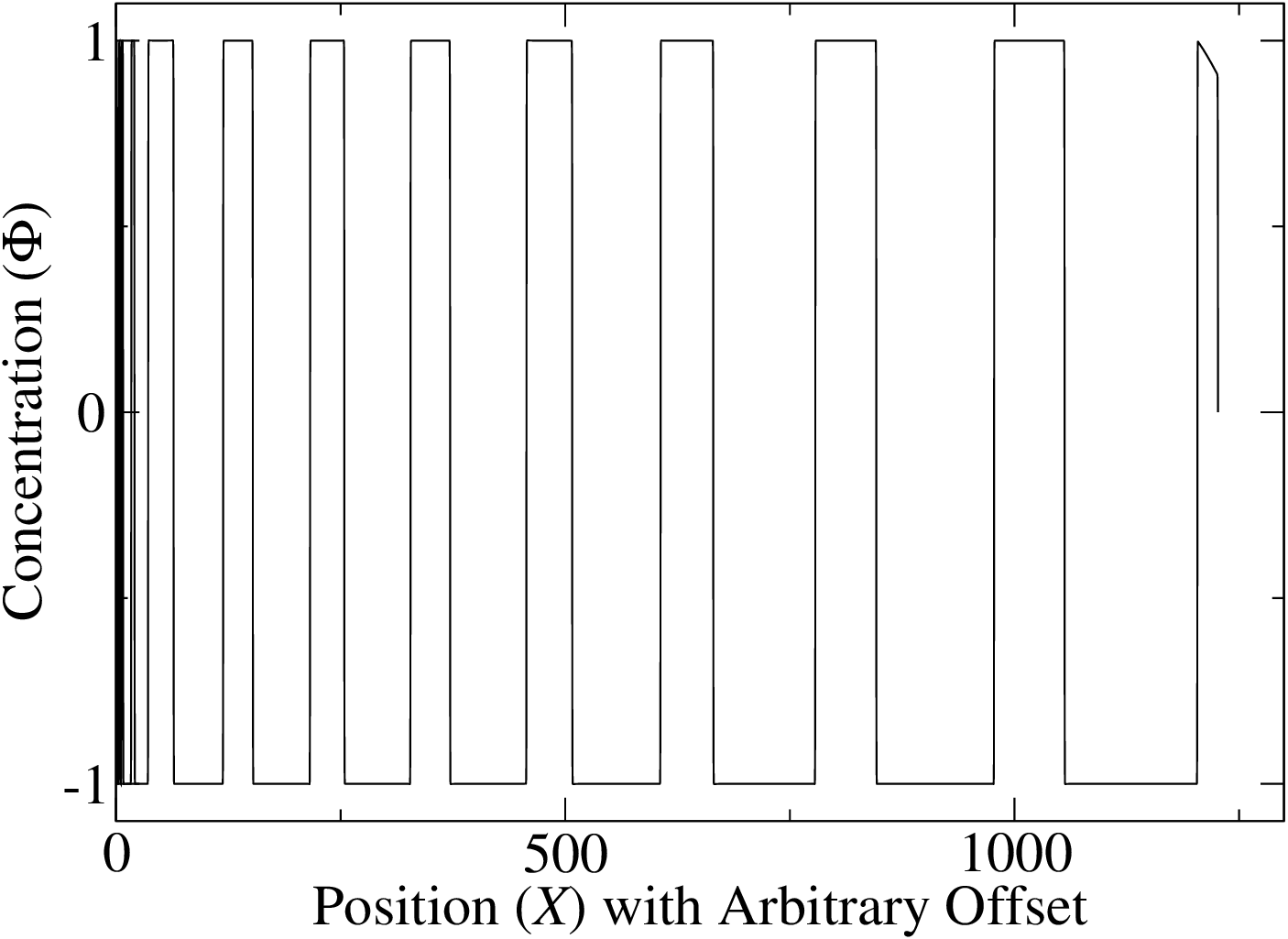}}
		\hfill
		\subfigure[]{\label{fig:off_fitting}\includegraphics[clip=true, scale=0.29]{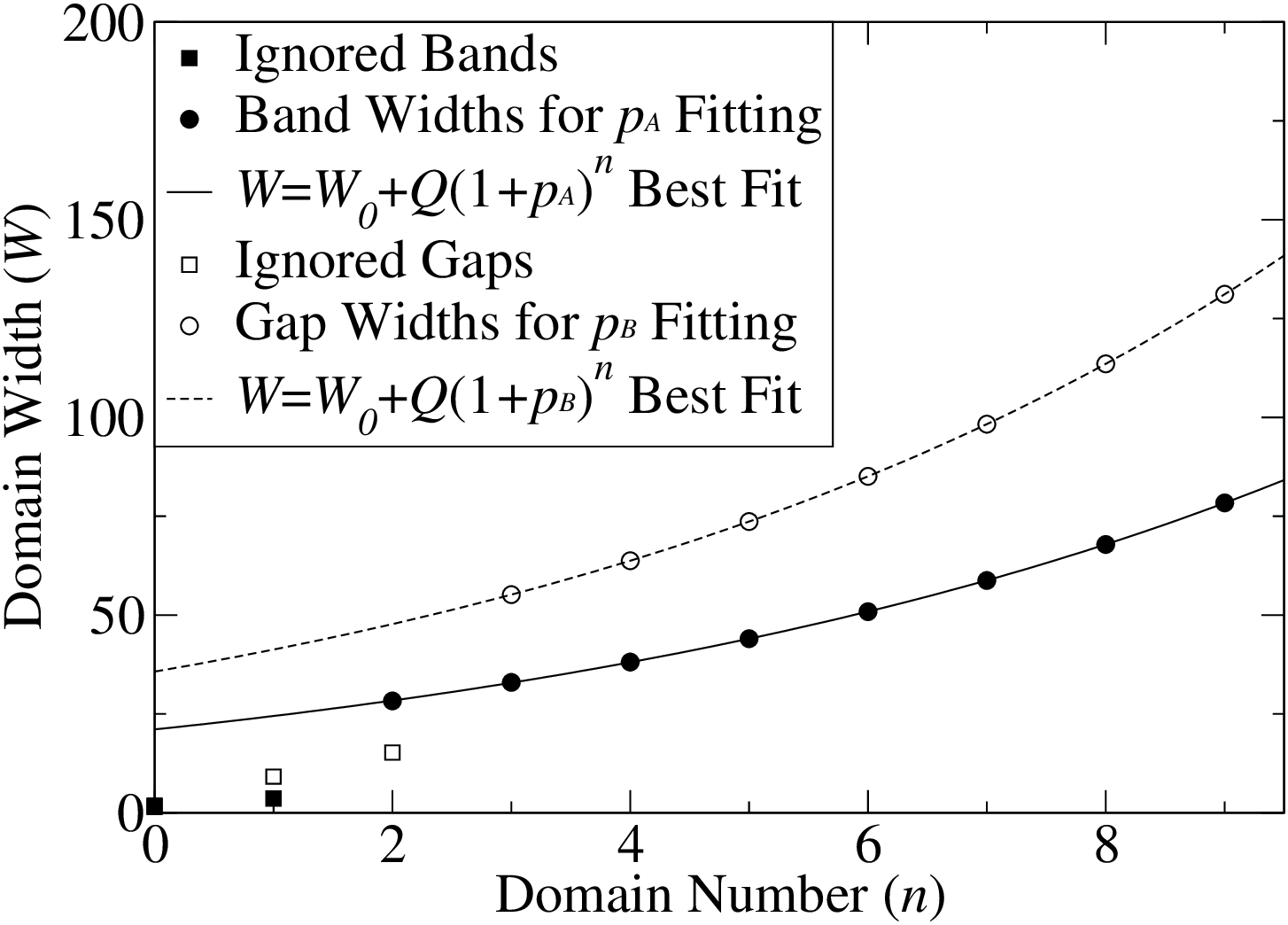}}
		\caption{Example concentration profile showing Liesegang patterns formed by an enslaved front moving through material with an initial concentration $\Phi_{in}=-0.3$, and a front speed scaling constant $C=0.5$.
			The profile is shown in subfigure (a) at simulation completion.
			Subfigure (b) shows fitting the band ($\mathcal{A}$-type) and gap ($\mathcal{B}$-type) widths separately to find $p(C)$ for initial material consisting of a majority $\mathcal{B}$-type.
			Fitting parameters for the bands are $W_0=-1.41736$, $Q=22.4956$, and $p_\mathcal{A}=0.151002$.
			Fitting parameters for the gaps are $W_0=-0.449771$, $Q=36.1552$, and $p_\mathcal{B}=0.15435$.
			This spacing coefficient is higher than for material which is initially equally mixed $\Phi_{in}=0$.  As can be seen in \figref{fig:p_theory}, the spacing coefficient for $C=0.5$ is $p\approx0.12$ in the case where $\Phi_{in}=0$.}
		\label{fig:off_sims}
	\end{center}
\end{figure}

In this paper we have shown that a model for an enslaved phase separation front which moves at diffusive speed $U=C/\sqrt{T}$ can be used to produce Liesegang patterns.  We have done this by deriving the Liesegang time, width, and spacing laws from the model.  Our analysis includes the determination of the Liesegang spacing coefficient $p$ as a function of the front speed parameter $C$.  In doing so we determined the values of $C$ where Liesegang patterns are produced, and verified this with direct numerical simulations of the model.

For this paper we have chosen the initial material composition to be symmetrically mixed.  This corresponds to $\Phi_{in}=0$ and allows us to use the analytical result in \eqnref{eqn:solution} to determine the Liesegang laws and their constants of proportionality.  This model, however, can generate Liesegang patterns for the range of initial concentrations between the spinodal concentrations.  An example is shown in \figref{fig:off_sims} of a Liesegang pattern for asymmetrically mixed initial material generated by this model.  The parameters used in this simulation were $C=0.5$ and $\Phi_{in}=-0.3$.  We expect that it will be possible to determine a spacing law for off-critical mixtures, but that is outside the scope of this paper.  Additionally, the production of Liesegang patterns by enslaved fronts which move into material with non-zero mixed mobility should be considered.  We expect the results to be qualitatively similar, but we are not sure than this case will be amenable to an analytical treatment.

\section*{Acknowledgments}
We would like to think S. May and G. Kaehler for helpful discussions.  E.F. acknowledges support by the National Science Foundation under grant DMR-0513393.

\newpage
\bibliographystyle{plain}

\end{document}